\documentclass{svjour3}                     % onecolumn (standard format)
\smartqed  % flush right qed marks, e.g. at end of proof
\usepackage{amssymb}
\usepackage{graphicx}
\usepackage{natbib}
\bibpunct{(}{)}{;}{a}{}{,}
\usepackage{epsfig}

\newcommand{\be}{\begin{equation}}
\newcommand{\ee}{\end{equation}}
\newcommand{\beq}{\begin{eqnarray}}
\newcommand{\eeq}{\end{eqnarray}}

\newcommand{\ion}[2]{#1\,{\sc{#2}}}
\journalname{SSRv}

\begin{document}\title{Metal enrichment processes}

\author{S. Schindler \and
  A. Diaferio 
}

\institute{S. Schindler \at
             Institute for Astro- and Particle Physics,
                  University of Innsbruck, Technikerstr. 25, 6020 Innsbruck, Austria\\
                  \email{Sabine.Schindler@uibk.ac.at}
\and
A. Diaferio \at
              Dipartimento di Fisica Generale ``Amedeo Avogadro'', Universit\`a degli Studi di Torino,
                  Via P. Giuria 1, I-10125, Torino, Italy \\
Istituto Nazionale di Fisica Nucleare (INFN), Sezione di Torino, Via P. Giuria 1, I-10125, Torino, Italy 
}
\date{Received: 12 November 2007 ; Accepted: 7 December 2007 }

\maketitle

\begin{abstract}
There are many processes that can transport gas from the
galaxies to their environment and enrich the environment in this way
with metals. These metal enrichment processes have a large influence
on the evolution of both the galaxies and 
their environment. Various processes can contribute to
the gas transfer: ram-pressure stripping, galactic
winds, AGN outflows, galaxy-galaxy interactions and others. We review
their observational evidence, corresponding simulations, their
efficiencies, and their time scales as
far as they are known to date. It seems that
all processes can contribute to the enrichment. There is not a single
process that always dominates the enrichment, because the efficiencies
of the processes vary strongly with galaxy and environmental properties.
\end{abstract}

\section{Introduction}
\label{Introduction}

The gas between the galaxies in a cluster - the Intra-Cluster Medium (ICM) - does not
only contain primordial 
elements, but also a considerable amount of heavy elements like Fe, Si,
S, or O (see 
\citealt{Werner2008} - Chapter 16, this volume) resulting in metallicities around 0.5 in
Solar units and sometimes even higher values.
A large fraction ($\approx 15-20$~\%) of the total mass of a cluster is in the ICM,
whereas the galaxies contribute a substantially smaller fraction ($3-5$~\%),
and the rest is dark matter. It follows that there is more mass in metals
in the ICM than in all the galaxies of a cluster.
This means that
 a lot of metals must have been transported from the galaxies into the ICM. This gas
transfer affects the evolution of galaxies and of galaxy
clusters. When galaxies lose their gas, the star formation rate
decreases and consequently the properties of the galaxies
change. Depending on the time and the efficiency of the gas
removal the evolution of the galaxies is more or less affected. Therefore
it is important to know when, where and how the gas transport takes place.

Various processes are discussed that can contribute to the metal
enrichment - some depend only on internal properties of the
galaxies, others on the environment or the combination of both. We review
here several enrichment processes: ram-pressure stripping, galactic
winds, AGN outflows, galaxy-galaxy interactions and the effect of an
intra-cluster stellar population.
Please note that this list is certainly not complete and further
processes might also contribute a small fraction to the metal
enrichment of the ICM. Furthermore, some processes influence each
other, which makes the picture even more complicated.

For several of the processes not only
observational evidence exists, but also numerical simulations have
been performed. We review here both aspects.

\section{Ram-pressure stripping}
\label{rps}

A galaxy passing through the ICM feels an external pressure. This
pressure depends on the ICM density $\rho_{\rm ICM}$  and the relative
velocity $v_{\rm rel}$ of the galaxy and the ICM. \citet{Gunn1972}
suggested this process already many years ago. They also gave a
frequently used prescription for the radius  $r$ beyond which the gas of a
galaxy is stripped, depending on the ram pressure and the galactic gravitational
restoring force. The
implicit condition on $r$ reads

\begin{equation}
p_{\rm ram} = \rho_{\rm ICM}v^2_{\rm rel} > 2\pi G\sigma_{\rm star}(r) \sigma_{\rm gas}(r)
\end{equation}
with $p_{\rm ram}$ being the ram pressure, $G$ the gravitational constant,
$\sigma_{\rm star}$ the stellar surface density, $\sigma_{\rm gas}$ the surface
mass density of the galactic gas.

\subsection{Observations}
\label{rps_obs}

Nowadays the process of ram-pressure stripping receives more and more
attention. There is now much observational
evidence of stripped galaxies. In the Virgo cluster several
examples of spiral galaxies affected by ram-pressure stripping have been found
by \ion{H}{i} observations \citep[see Fig.~1]{Cayatte1990,Veilleux1999,Vollmer1999,Vollmer2003,Kenney2004,Vollmer2004a,Vollmer2004b,Koopmann2004,Crowl2005}. Furthermore in Virgo
elliptical galaxies stripping features have been discovered (e.g.
\citealt{Rangarajan1995,Lucero2005,Machacek2006a}). Also in Coma and
other clusters and groups evidence for ram-pressure stripping has been found
\citep{Bravo-Alfaro2000,Bravo-Alfaro2001,Kemp2005,Rasmussen2006,Levy2007}. Deficiency of \ion{H}{i} has been
reported as evidence for ram-pressure stripping \citep{Vollmer2007}. It is even possible that the \ion{H}{i}
plume of the galaxy NGC~4388, that extends to more than 100 kpc, is
a ram-pressure stripping feature \citep{Oosterloo2005}. In the galaxy NGC~7619  stripping features have been found 
showing a high metallicity in the gas tail behind the galaxy \citep{Kim2007}. The galaxy ESO 137-001 in the cluster Abell 3627 shows a
long tail with several star formation regions suggesting that
ram-pressure triggers star formation not only within the galaxy but
also in the stripped material \citep{Sun2008}.
Recently ram-pressure features have also been found in 
distant clusters \citep{Cortese2007}.

\begin{figure}    %%%%%%%%%%%%%%%%%% FIGURE 1
\begin{center}
\psfig{file=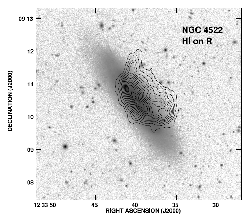,width=8.cm,clip=}
\vskip .1cm
\psfig{file=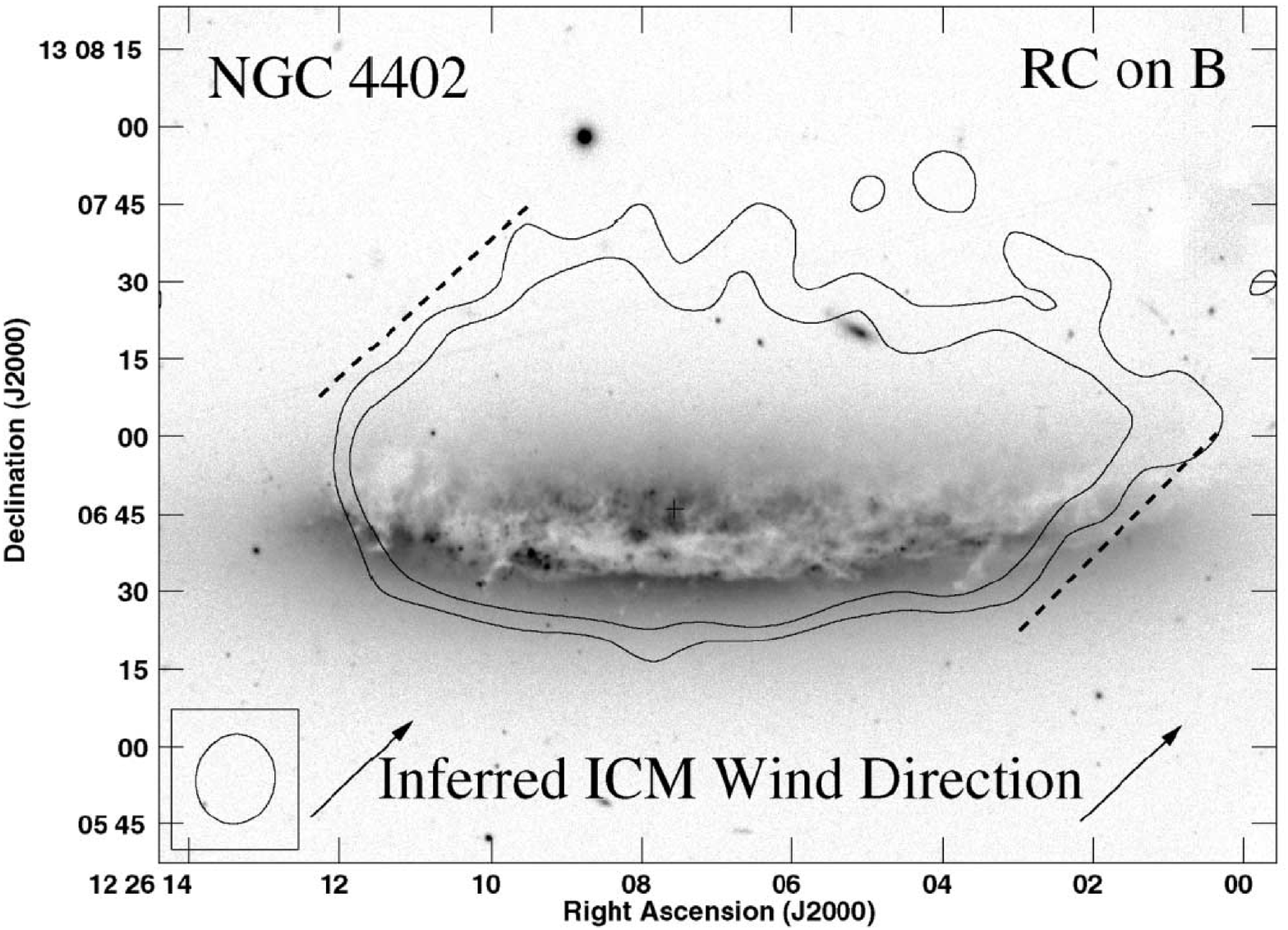,width=8.cm,clip=}
\caption{Two examples of galaxies affected by ram-pressure stripping in the Virgo
  cluster. Top: \ion{H}{i} gas (contours) and stellar light (grey
  scale, R-band) of
  NGC~4522. While the stellar distribution looks undisturbed the 
  gas is bent back due to ram-pressure stripping (1.4 GHz radio
  continuum contours on B-band, from \protect\citealt{Kenney2004}). Bottom: Similar features in NGC~4402 (from \protect\citealt{Crowl2005}).
}
\end{center}
\end{figure}

For a review on ram-pressure stripping and \ion{H}{i} deficiency see \citet{vanGorkom2003}.

\subsection{Simulations}

As ram-pressure stripping is such a common process, there are many
simulations in which the stripping process of galaxies was calculated for
different types of galaxies: spirals, ellipticals and dwarfs \citep[see Fig.~2]{Abadi1999,Quilis2000,Mori2000,Toniazzo2001,Schulz2001,Vollmer2001,Hidaka2002,Bekki2003,Otmianowska-Mazur2003,Acreman2003,Marcolini2003,Roediger2005,Roediger2006a,Roediger2006b,Mayer2006,Vollmer2006}. The simulations confirm that
the process is acting in the expected way. Starting from the outer
parts of the galaxy gas is stripped off. Part of this gas is not bound to
the galaxies anymore and left in a wide (fragmenting) tail behind the
galaxy. Recently even the increase of star formation in and behind the
galaxy caused by ram-pressure stripping has been found in simulations
\citep{Kronberger2008a,Kapferer2008}. \citet{Bruggen2008} found that more than half of the cluster galaxies have experienced
ram-pressure stripping and hence a considerable fraction of galaxies
in a cluster at the present epoch has suffered a
substantial gas loss.

\begin{figure}    %%%%%%%%%%%%%%%%%% FIGURE 2
\begin{center}
\psfig{file=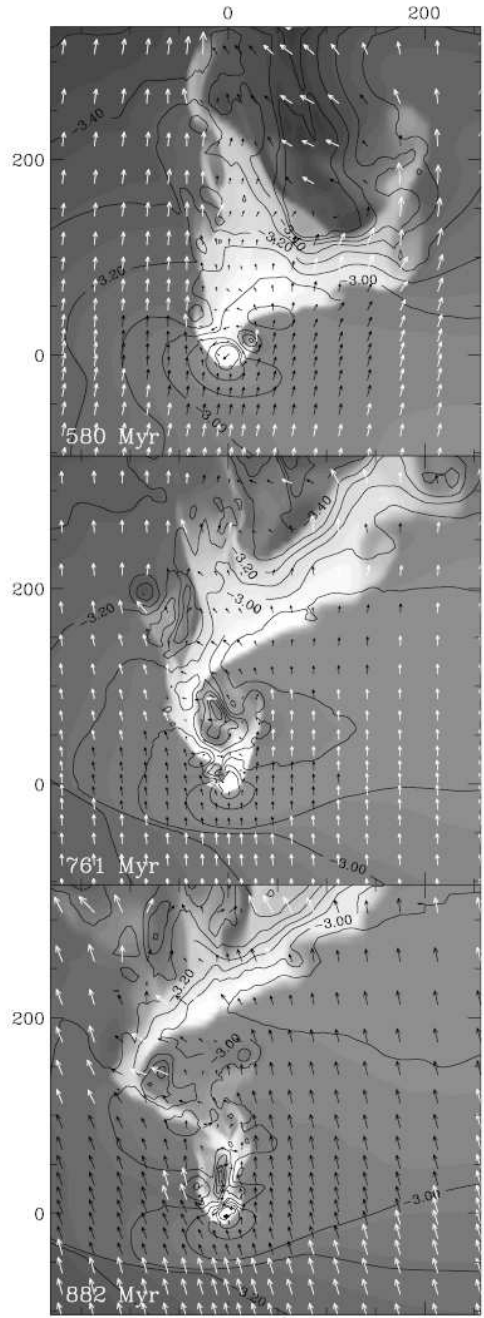,width=6.5cm,clip=}
\caption{Simulation of a galaxy in the process of ram-pressure
  stripping with gas density (grey 
  scale, ranging from $4\times 10^{-29}$ to $1 \times 10^{-27}$ g\,cm$^{-3}$), pressure (black contours, logarithmically spaced from
  ${\rm log}(p)=-3.6$ to ${\rm log}(p)=-2.6$ in units of $10^{-24}$ keV\,cm$^{-3}$) and
  velocity vectors (white when Mach number $> 1$, and black
  otherwise) at three 
  different times. A cut through a 3D simulations is shown
  with coordinates in kpc. Due to the ram pressure 
  of the ICM the galaxy loses more and more of its gas
(from \protect\citealt{Toniazzo2001}).
}
\end{center}
\end{figure}

An analytical model has been developed to describe ram-pressure
stripping for galaxies of different morphologies in different
environments \citep{Hester2006}. It
describes the stripping of a satellite galaxy's outer \ion{H}{i} disk and hot galactic halo.

It was tested with simulations whether the simple, widely used
criterion by \citet{Gunn1972} is a good estimate for the mass loss.
Generally it is found that the criterion is a good
estimate for the mass loss \citep{Roediger2007b,Kronberger2008a} when simulations and analytic estimates are compared for the
same conditions (see \citealt{Jachym2007} for a comparison with different conditions)  - 
a quite surprising result given the simple assumptions of the
criterion, that do not even take into account dark matter.

\section{Galactic winds}

Already many years ago galactic winds were suggested as a possible
gas transfer mechanism \citep{DeYoung1978}. Many supernova explosions provide
large amounts of thermal energy, which can drive an outflow from a
galaxy (see reviews by \citealt{Heckman2003} and \citealt{Veilleux2005}). A correlation between starburst galaxies and wind is well
established through the finding of hot gas around starburst galaxies
(e.g. \citealt{Dahlem1998}).

Spectacular examples of such winds are seen in the galaxies M~82  \citep{Lynds1963}
and NGC~253 \citep{Demoulin1970}.

The outflows consist of a complex
multi-phase medium of cool, warm and hot gas (see e.g. the {\sl Chandra}
observation of NGC~4631, \citealt{Wang2001}, Fig.~3).
The morphologies of the optical
emission-line gas and the X-ray emission as observed with {\sl Chandra} have
been found to be quite
similar \citep{Strickland2002,Cecil2002}. Such correlations can be used
to understand the interaction between the gas
in the bubbles and the interstellar medium (ISM). It was found that the accelerated ISM can reach high velocities of
several hundred km\,s$^{-1}$ \citep{Heckman2000,Rupke2002}.

\begin{figure}    %%%%%%%%%%%%%%%%%% FIGURE 3
\begin{center}
\psfig{file=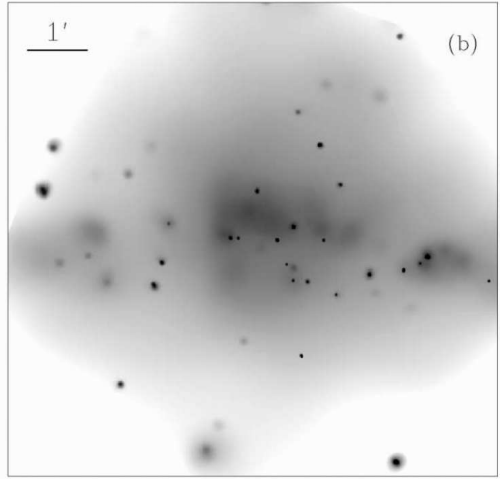,width=7.cm,clip=}
\caption{{\sl Chandra} observation of the edge-on spiral galaxy NGC~4631.
  It shows the presence of a giant diffuse X-ray emitting corona. The corona
  has a temperature of $(2-7) \times10^6$ K and extends as far as 8 kpc away
  from the galactic plane 
(from \protect\citealt{Wang2001}).
}
\end{center}
\end{figure}

With these winds also metals are
transported into the ICM. The amount
of metals depends on various galaxy parameters, like the total mass of
the galaxy or the disc scale length,
and on the environmental conditions: e.g. in the centre
of massive clusters the pressure of the ICM can suppress the winds
\citep{Kapferer2006}. This suppression typically takes place for
ICM pressures above $(0.7-1) \times 10^{12}$ dyne\,cm$^{-2}$.

\citet{Martin1999} gives an often used recipe for simulations: the mass
outflow rate $\dot M$ is proportional to the
star formation rate SFR:
\begin{equation}
\dot M\, =\, \epsilon\, {\rm SFR}
\end{equation}
with $ \epsilon$ being typically in the range of $1-3$. By comparing
different techniques \citet{Heckman2003} also finds 
that the outflow rate is of the order of the
star formation rate.  
The SFR can be estimated from observations,
e.g. from far-infrared luminosities $L_{\rm FIR}$
\begin{equation}
{{\rm SFR} \over 1\, {\rm M}_\odot {\rm yr}^{-1}} = {L_{\rm FIR} \over 5.8 \times 10^9 {\rm L}_\odot}
\end{equation}
\citep{Kennicutt1998}. Another way to estimate the SFR is to use the tight
relation between the SFR and the surface density of the gas $\sigma_{\rm gas}$
\begin{equation}
\Sigma_{\rm SFR} \propto \sigma_{\rm gas}^N
\end{equation}
\citep{Schmidt1959} with $\Sigma_{\rm SFR}$ being the surface density of the SFR
 and the index $N$ having
measured values between 1 and 2. Only at densities below a critical
threshold value the SFR is almost completely suppressed \citep{Kennicutt1989}. Alternatively the dynamical time $t_\ast$ can be included 
\begin{equation}
\Sigma_{\rm SFR} \propto {\Sigma_{\rm gas} \over t_\ast}
\end{equation}
with the dynamical time $t_\ast$ being the local orbital timescale of the disk
 \citep{Kennicutt1998}.
For hydrodynamic simulations this has been extended to include the
fraction of stars lost by supernova explosions
by \citet{Springel2003}
\begin{equation}
{{\rm d}\rho_{\ast} \over {\rm d}t} = (1-\beta) {{\rm d}\rho_{\rm c} \over {\rm d}t}
\end{equation}
with $\rho_{\ast}$ being the density of stars, $\rho_{\rm c}$ being the cold gas density in the disk and $\beta$ being the
fraction of stars lost by supernova explosions. For a typical initial
mass function and a mass threshold of 8 $M_\odot$ for the supernovae a $\beta=0.1$ is
used. These are of course  
only statistical estimates.

Other attempts to quantify
the outflow rate take into account physical parameters like those describing 
the galaxy's gravitational potential and the effect of cosmic rays \citep{Breitschwerdt1991}.
Using  the Bernoulli equation, \citet{Kronberger2008b} recently derived an analytic
approximation for the mass loss due to thermally driven galactic
winds. The mass 
loss per unit area at a given position of the galactic disc reads
\begin{equation}
\dot{M}=\rho_0 u_0=\rho_0\sqrt{v_{\rm esc}^2+2\Phi_0-\frac{5}{\gamma}c_0^2},
\end{equation}
with $\rho_0$ being the gas mass density, 
$u_0$ the bulk velocity of the gas, 
$\Phi_0$ the gravitational
potential, $c_0$ the sound speed (all four quantities at the given position),
$v_{\rm esc}$ the escape velocity, 
and $\gamma$ the adiabatic index of the thermal plasma. 
Hydrodynamic simulations of outflows have also been performed
\citep{Tenorio-Tagle1998,Strickland2000}.

Starbursts with subsequent winds can also be caused by cluster mergers
\citep{Ferrari2003,Ferrari2005,Ferrari2006}, because in such mergers the
gas is compressed and shock waves and cold fronts, which trigger star
formation, are produced 
\citep{Evrard1991,Caldwell1993,Wang1997,Owen1999,Moss2000,Bekki2003}.

\section{Galaxy-galaxy interaction}

Another possible mechanism for removing material -- gas and stars --
from galaxies is the interaction between the galaxies (e.g.
\citealt{Clemens2000,Mihos2005}, see Figs.~4, 5 and 6). While the direct
stripping effect is mostly not very efficient in clusters due to
the short interaction times, the close passage of another galaxy
(sometimes called galaxy harassment)
can trigger a star burst \citep{Barnes1992,Moore1996,Bekki1999}, which subsequently can lead to a galactic wind
\citep{Kapferer2005}. But there can be a competing effect: the
ISM might be stripped off immediately by ram pressure
\citep{Fujita1999,Heinz2003a} and hence the star
formation rate could drop. In any case ISM would be removed from
the galaxies.

\begin{figure}    %%%%%%%%%%%%%%%%%% FIGURE 4
\begin{center}
\psfig{file=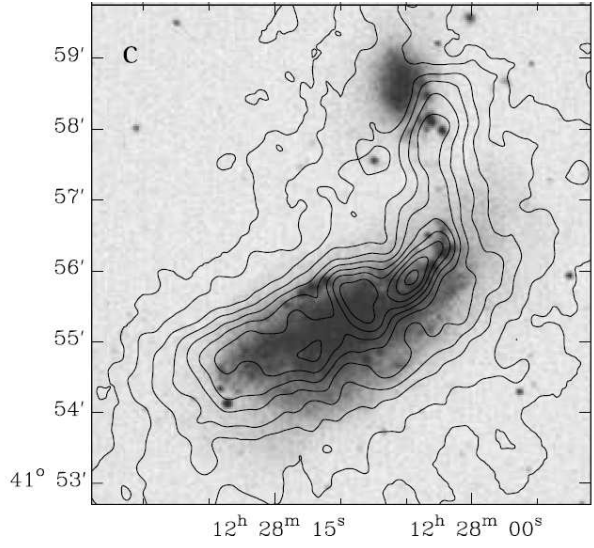,width=8.cm,clip=}
\caption{Image of the  interacting galaxies NGC~4490 / NGC~4485 in \ion{H}{i}
  (contours) and optical R band (grey scale). Some of the gas is lost
  due to the interaction of the galaxies (from \protect\citealt{Clemens2000}).
}
\end{center}
\end{figure}

Simulations of interactions between galaxies containing an AGN show a complex interplay
between star formation and the activity of the AGN itself \citep{Springel2005}.

\begin{figure}    %%%%%%%%%%%%%%%%%% FIGURE 5
\begin{center}
\psfig{file=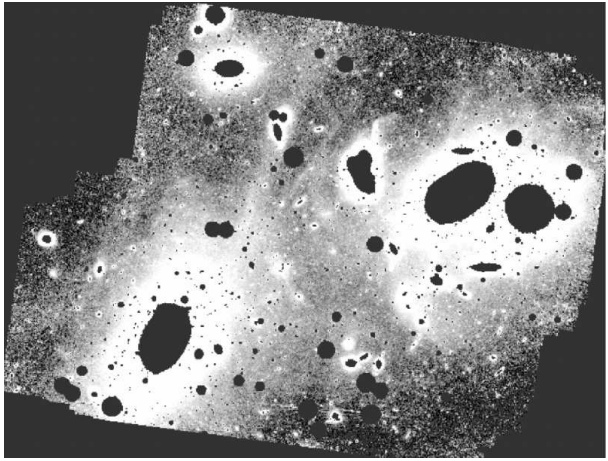,width=8.cm,clip=}
\caption{Very deep observation of the core of the Virgo
  cluster. Diffuse light is visible between the galaxies which results
  from stars that have been expelled from the galaxies due to
  interactions between them (from \protect\citealt{Mihos2005}).
}
\end{center}
\end{figure}

In order to estimate how likely such interaction events are the
number of encounters and mergers needs to be assessed.
The number of close encounters that a galaxy experiences within
$t_{\rm H} = 10^{10}$ years was estimated by \citet{Gnedin2003} in the following
way. He assumed a galaxy of size $R_{\rm g} = 10$ kpc, a virialised cluster
with a one-dimensional velocity dispersion of $\sigma_{\rm cl} = 1000$~km\,s$^{-1}$, a virial radius of $R_{\rm cl} = 1$ Mpc and $N_{\rm g} =
1000$ galaxies uniformly distributed within this radius. With a
relative velocity of $\sqrt2\sigma_{\rm cl}$ and neglecting the
gravitational focussing, he finds
\begin{equation}
N_{\rm enc} \approx {{N_{\rm g}} \over {(4\pi/3)R_{\rm cl}^3}} \pi R_{\rm g}^2\sqrt2
\sigma_{\rm cl}t_{\rm H} \approx 1,
\end{equation}
i.e. a galaxy is expected to encounter one other galaxy over the
course of its evolution. Even though the assumption is very
simplifying one sees that an encounter is a relatively
frequent event in a cluster. In contrast to this, \citet{Gnedin2003} estimated
for the probability to merge with another
galaxy 
\begin{equation}
P_{\rm mer} \approx N_{\rm enc} \biggl({\sigma_{\rm g}\over\sigma_{\rm cl}}\biggr) ^4 \approx 10^{-3} 
\end{equation}
with a galactic velocity dispersion $\sigma_{\rm g} = 200~$km\,s$^{-1}$. Hence an
actual merger is an unlikely event.

Tidal interactions and merging between galaxies
are highly non-linear phenomena that can be partly handled
analytically (e.g., \citealt[chapter 7]{Binney1987}). Numerical simulations, however, are required for accurate
estimates of mass loss rates and the morphological modification of
galaxies.

\begin{figure}    %%%%%%%%%%%%%%%%%% FIGURE 6
\begin{center}
\psfig{file=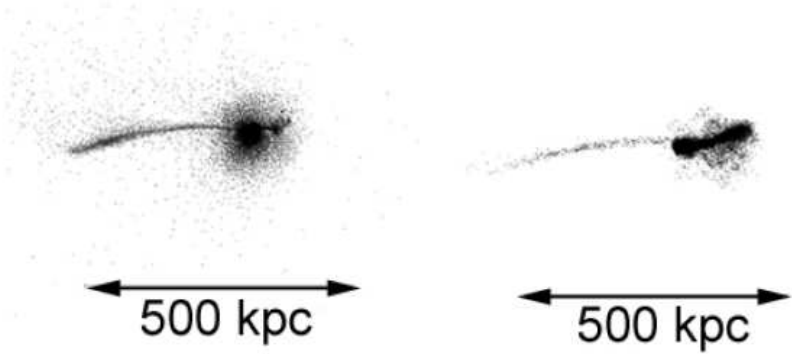,width=12.cm,clip=}
\caption{Simulation of the interaction between two galaxies:
  distribution of gas (left) and stars (right). Due to the interaction
  gas and stars can be expelled out to distances of hundreds of kpc
(from \protect\citealt{Kapferer2005}).
}
\end{center}
\end{figure}

\section{AGN outflows}

We discuss two types of outflows from AGN: jets and winds-like outflows.
There is much observational evidence for AGN jets
interacting with the ICM -- not only radio jets but also cavities in
the ICM found in X-rays (e.g.  \citealt{Blanton2001,McNamara2001,Schindler2001,Heinz2002,Choi2004,Fabian2006,McNamara2007}), in
which the pressure of the relativistic particles of the jet has pushed away
the ICM. The jets consisting of relativistic particles can entrain
some of the surrounding  metal-rich gas \citep{DeYoung1986}.

As the jet-ICM interaction can have an effect on both the
energetics and the metal enrichment of the ICM, several groups have
calculated this process. Many simulations for the
energy transfer have been performed \citep{Zhang1999,Churazov2001,Bruggen2002,Nulsen2002,Krause2003,Heinz2003b,Beall2004,Beall2006,DallaVecchia2004,Zanni2005,Sijacki2006,Heinz2006} while only few
have attempted to calculate the metal enrichment due to the entrainment by
jets \citep{Heath2007,Moll2007}. These simulations found
that jets can both heat and enrich the ICM considerably. Another type of
simulations calculated the metallicity distribution due to
bubble-induced motions coming from a single AGN in the cluster centre
\citep{Roediger2007a}. It was found that in this case the metallicity
distribution is 
very elongated along the direction of the motion of the bubbles.

Also for wind-like outflows there is some observational
evidence. Blue-shifted absorption lines have been observed in UV and
X-rays \citep{Crenshaw2003}. Also from X-ray imaging evidence for
nuclear outflows has been found \citep{Machacek2006b}.
There are hints for a high
metallicity of a few times solar \citep{Hamann2001,Hasinger2002}, for high velocities of several thousands or several ten
thousands of km\,s$^{-1}$ \citep{Chartas2002,Chartas2003,Pounds2003a,Pounds2003b,Reeves2003,OBrian2005,Dasgupta2005,Gabel2006} and for considerable mass outflow
rates \citep{Crenshaw2003,Veilleux2005}.
The outflows can be quite strong, e.g. several $10^9$~M$_{\odot}$ with kinetic
energies around $10^{60}$ erg expelled over the AGN live time of $10^7$ years
as estimated from spectroscopic studies \citep{Nesvadba2006}. 

In some galaxies the winds are not only driven by repeated supernova
explosions but also the AGN are contributing to the energy necessary for
the wind (see Sect. 3).

\section{Intra-cluster stellar population}

There is increasing evidence for a population of stars in the space between
the galaxies in a cluster \citep{Bernstein1995,Gonzalez2000,Gonzalez2005,Gerhard2002,Gerhard2005,Gal-Yam2003,Arnaboldi2004,Cortese2004,Feldmeier2004,Ryan-Weber2004,Adami2005,Zibetti2005,Krick2007}. Depending on the mass of
the cluster the fraction of intra-cluster stars (= ratio of number of stars between
galaxies to total number of stars)  can be as high as
$10-50$~\% with the higher fraction being in more massive clusters.
This stellar population can originate from
stripping of stars from galaxies due to tidal interaction \citep{Cypriano2006}, can be expelled during mergers and the formation of massive
galaxies \citep[Fig.~6]{Murante2007,Kapferer2005} or can have multiple origins \citep{Williams2007}. Simulations show that intra-cluster stars should be
ubiquitous in galaxy clusters \citep{Willman2004} and their numbers should
generally increase with time \citep{Rudick2006}. A link between the
growth of the brightest cluster galaxy and the intra-cluster light
was reported by \citet{Zibetti2005}.

When these stars
explode as supernovae (mainly type Ia, as it takes a while for the stars to
travel away from the galaxies) they can enrich the ICM very efficiently because
there is no ISM pressure around them to confine the metals
\citep{Domainko2004,Zaritzsky2004,Lin2004,Dado2007}.

Considerably more frequent than supernova Ia explosions are their progenitors -
the recurrent  novae. With about $10^{-4}$~M$_{\odot}$ outflow per nova event
and typically super-solar abundances (up to ten times Solar,
\citealt{Gehrz1998}), novae could also contribute to the metal enrichment of the
ICM if they have been expelled previously from the galaxies.

A fraction of the AGB stars are also expected to be between the
galaxies. These stars have a considerable mass loss with metallicities of
about Solar abundances with slightly enhanced abundances of CNO elements
\citep{Wheeler1989,Zijlstra2006,vandenHoek1997,Busso2001,Nordstrom2003}. 
As the ratio of planetary nebulae (PNe) to AGB stars is well studied in 
statistical studies of PNe in the ICM  \citep{Feldmeier1998,Theuns1997,Arnaboldi2003} this ratio may be used to estimate the number of AGB stars.

In conclusion the population of stars should also be considered for
the enrichment processes in the ICM -- even far away from galaxies.

\section{Which of these processes are important for the ICM enrichment?}

Several groups have addressed this question already many years
ago. \citet{David1991} proposed the first models taking into
account the effects of galactic winds on the ICM enrichment. They
found that the results depend sensitively on their input parameters:
the initial mass function, the adopted supernova rate and the primordial mass fraction of
the ICM. 

The first 3D simulations calculating the full gas dynamics and the
effects of winds on cluster scales were performed by \citet{Metzler1994,Metzler1997}. They concluded that winds can account for the observed
metal abundances in the ICM, but they found strong metallicity
gradients (almost a factor of ten between cluster centre and virial
radius) which are not in agreement with observations. \citet{Gnedin1998}
took into account not only galactic winds, but also galaxy-galaxy
interactions and concluded that most of the metals are ejected by
galaxy mergers. In contrast to this result \citet{Aguirre2001} found
that galaxy-galaxy interactions and ram-pressure stripping are of
minor importance while galactic winds dominate the metal enrichment of
the ICM. 

That these early results disagree so much is probably due to
the large range of scales that is involved. On the one hand the whole
cluster with its infall region has to be simulated´, on the other hand
processes within galaxies or even within the active core of a galaxy
are important. It is not possible to calculate all of this accurately
in one type of simulation and therefore new methods have been
developed. 

Recently several simulations for the
 enrichment have been performed. They calculate the exact composition and evolution of the
 ISM by varying the initial mass function and the yields of supernova explosions (see
 \citealt{Borgani2008} - Chapter 18, this volume), but it is not distinguished by which
 process the enriched gas is transported into the ICM. Specially for
 the transport processes a new simulation method has been developed, in
 which N-body/hydro\-dynamic simulations with mesh refinement
 including a 
semi-analytical method have been combined with separate descriptions
of the various enrichment processes, which can be switched on and off
individually \citep{Schindler2005}. 

The results obtained with this method show an inhomogeneous
distribution of the metals independent of the enrichment processes
\citep[see Fig.~7]{Schindler2005,Domainko2006,Kapferer2006,Moll2007}. These results are in very good
agreement with the observed metallicity maps (see \citealt{Werner2008} -
Chapter 16, this
volume). The gas lost by the galaxies is obviously not mixed
immediately with the ICM. There are usually several maxima visible in
the metallicity distribution, which are not necessarily associated
with the cluster centre. The maxima are typically at places where galaxies
just have lost a lot of gas to ICM of low density, mostly due to 
star bursts. The metallicities vary locally between 0 and 4 times
Solar.

\begin{figure}    %%%%%%%%%%%%%%%%%% FIGURE
\begin{center}
\psfig{file=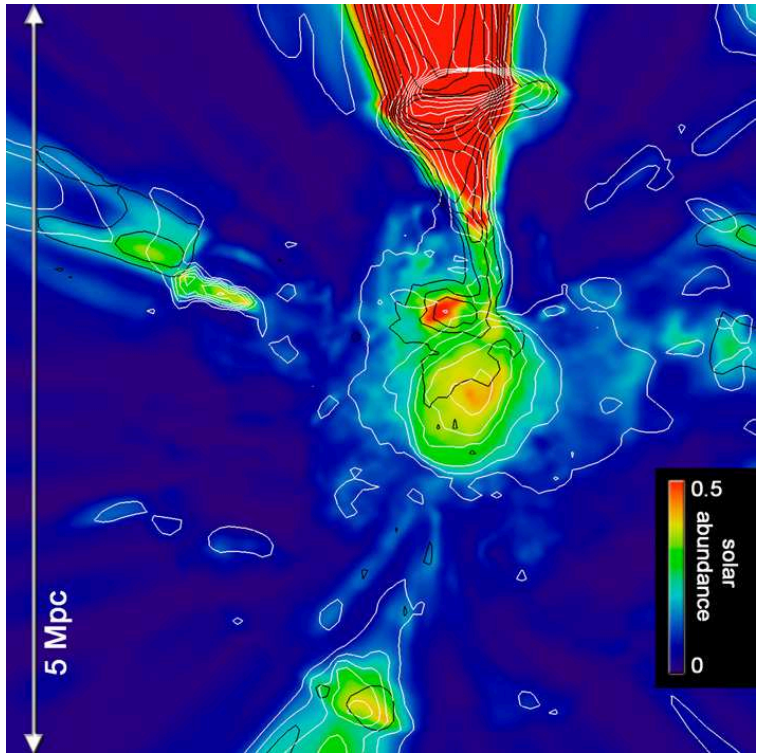,width=10.cm,clip=}
\caption{Simulated metallicity map, i.e. an X-ray emission weighted,
  projected metal distribution. The high metallicity region at the top
  is caused by a group of galaxies with recent starburst. Overlaid
  contours indicate the origin of the metals: ram-pressure stripping
  (white) and galactic winds (black) (adopted from \protect\citealt{Kapferer2007a}).
}
\end{center}
\end{figure}

A detailed comparison between the two enrichment mechanisms - winds and
ram-pressure stripping - revealed that these two processes yield
different metal distributions  (see Fig.~7) and a different time dependence of the
enrichment \citep{Kapferer2007a,Rasia2007}. The
ram-pressure stripped gas is more centrally concentrated. The reason
for this is that the ICM density as well as the galaxies´
velocities are larger in the cluster centre, so that ram-pressure
stripping is very efficient there. Galactic winds, however, can be
suppressed by the high pressure of the ICM in the centre \citep{Kapferer2006}, so that in massive clusters galactic winds do hardly
contribute to the central enrichment. The resulting radial metal profiles
are correspondingly relatively flat for galactic winds and steep for
ram-pressure stripping. When both processes are taken into account they
are in good agreement with the observed profiles (see also \citealt{Borgani2008} - Chapter 18, this volume).

The time scales for the enrichment are also different for the two
processes \citep{Kapferer2007a}. The mass loss of galaxies due to
winds is larger at high redshifts. Between redshifts 2 and 1
ram-pressure stripping becomes more important for the mass loss and
it is
by far more efficient at low redshift (see Fig.~8). The reason is that
on the one hand galactic winds become weaker because the star
formation rate decreases and on the other hand ram-pressure stripping
becomes stronger because clusters with ICM have formed/are forming,
which is interacting with the galaxies. In total the
mass loss due to ram-pressure stripping is usually larger than the
mass loss due to winds, in some cases up to a factor of three.

\begin{figure}    %%%%%%%%%%%%%%%%%% FIGURE 8
\begin{center}
\psfig{file=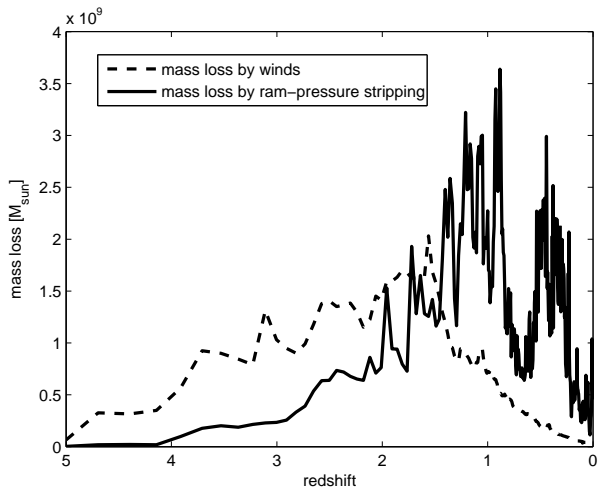,width=8.cm,clip=}
\caption{Mass loss of the galaxies in a simulation taking into
  account mass loss due to galactic winds (dashed line) and mass loss
  due to ram-pressure stripping (solid line) at different redshifts
  (adopted from \protect\citealt{Kapferer2007a}).
}
\end{center}
\end{figure}

Generally it is very hard to provide numbers for the relative
efficiencies of the various processes as the efficiencies depend
strongly on the properties of the clusters. In a massive or in a
merger cluster, for example, ram-pressure stripping is very
efficient. 

The simulated metallicities can be converted to artificial X-ray
metallicities, metallicity profiles, metallicity maps and metallicity
evolution. There is in general a good agreement between these quantities
derived from simulation and observation \citep{Kapferer2007b}. The metallicity
values are in the right range and  the spatial distribution and
the evolution are in good agreement with the observations. Also the
evolution of the metallicity since $z=1$ found in observations \citep{Balestra2007,Maughan2008} can be reproduced by the simulations.
 Of course
there is a large scatter in all these quantities, because they vary
very much from cluster to cluster both in simulations and observations. 

Summarising, from the comparison of observations with simulations it
seems clear that several processes are involved in the metal
enrichment and none of them can be ruled out immediately as being not
efficient enough. 
The processes can also influence each other (e.g. AGN outflows can
enhance an existing galactic 
wind or one process can suppress another one). Obviously the
interaction between galaxies and the ICM is a very complex 
issue. In order to know what is really going on at the transition
between galaxies and ICM many more 
observations and simulations are needed.

\begin{acknowledgements}
The authors thank ISSI (Bern) for support of the team ``Non-virialized
X-ray components in clusters of galaxies''. We thank Wolfgang Kapferer
und Thomas Kronberger for useful discussions.
The authors acknowledge financial support by the
Austrian Science Foundation (FWF) through grants P18523-N16 and
    P19300-N16, by the Tiroler Wissenschaftsfonds and through the
    UniInfrastrukturprogramm 2005/06 by the BMWF. Partial support from
the PRIN2006 grant ``Costituenti fondamentali dell'Universo'' of
the Italian Ministry of University and Scientific Research
and from the INFN grant PD51 is also gratefully acknowledged.
\end{acknowledgements}

\end{document}